\documentclass[twocolumn,amsmath,amssymb,pra,superscriptaddress]{revtex4-1}
\usepackage{graphicx}
\usepackage{dcolumn}
\usepackage{bm}
\usepackage{subfigure}
\usepackage{booktabs}
\usepackage{color}
\newcounter{one}
\def\bra#1{\mbox{\boldmath $#1$}^{\top}}
\def\ket#1{\mbox{\boldmath $#1$}}
\newcommand{\bracket}[1]{\left\langle #1 \right\rangle}

\newcommand{\affA}{
Department of Mathematical and Computing Science,
Tokyo Institute of Technology, 4259-G5-22, Nagatsuta-cho, Midori-ku, Yokohama, Kanagawa 226-8502, Japan
}

\begin{document}

\title{\textbf{Detectability thresholds of general modular graphs}}

\author{Tatsuro Kawamoto}
\affiliation{\affA}
\author{Yoshiyuki Kabashima}
\affiliation{\affA}

\begin{abstract}
We investigate the detectability thresholds of various modular structures in the stochastic block model. Our analysis reveals how the detectability threshold is related to the details of the modular pattern, including the hierarchy of the clusters. We show that certain planted structures are impossible to infer regardless of their fuzziness. 
\end{abstract}
 
\maketitle

\section{Introduction}\label{Introduction}
Motivated by needs in data-driven science, a number of frameworks and algorithms for modular structure detection have been proposed in several fields in the last few decades \cite{GirvanNewman2002,ShiMalik2000,Leskovec2009,Fortunato201075,Leger2013}. Correspondingly, theoretical and experimental analyses of statistical significance of results are thus the subject of significant research interest. For example, although an algorithm suggests the partition of a graph following the application of some optimization process, if the graph is a typical instance of a uniform random graph, it is doubtful whether the effected partition contains any useful information in practice. Moreover, even when the graph is generated from a model with some planted structure, it may be indistinguishable from a uniform random graph if the planted structure is too fuzzy. 

It is a challenging problem in general, and the basic strategy to solve it involves investigating the conditions whereby we can retrieve the planted structure for a specified random graph ensemble. 
To this end, the so-called stochastic block model \cite{holland1983stochastic}, which we explain in detail below, is often considered. This random graph model has controllable noise strength $\epsilon$, i.e., $\epsilon = 0$ represents a graph that clearly realizes the planted structure, and $\epsilon = 1$ represents a uniform random graph. Above a certain critical value $\epsilon^{\ast}$, an algorithm cannot retrieve the planted structure better than chance. 
This critical value is called the \textit{detectability threshold}, and a large number of studies have been devoted to it \cite{Reichardt2008,Nadakuditi2012,Krzakala2013,KawamotoKabashimaPRE2015,KawamotoKabashimaEPL2015,Decelle2011,Decelle2011a,Radicchi2013,Radicchi2014,Hu2012,Ronhovde2012,VerSteeg2014,ZhangPRE2014,Ghasemian2016} for sparse graphs, including rigorous treatments \cite{Mossel2014,Massoulie2014,banks2016information}. 
Besides the distinguishability from a uniform random graph, the exact recovery in dense graphs has also been studied \cite{Condon2001,Onsjo2006,BickelChenPNAS2009,Rohe2011,yun2014community,AbbeFOCS2015,AbbeNIPS2015}.

Nevertheless, a large portion \cite{FootnoteAbbe} of the research focuses on the community structure (assortative structure) and the disassortative structure. In this paper, we investigate the detectability threshold of more general structures. We show that according to the linear stability analysis of \textit{belief propagation} (BP), the detectability threshold varies depending on the details of the modular structure.

\section{Stochastic block model}
The stochastic block model is a random graph model with a planted modular structure: the graph of $N$ vertices consists of $q$ clusters, each of which of size $\gamma_{\sigma}N$ ($\sigma \in \{1, \dots, q\}$), and every pair of vertices is connected independently and randomly according to its cluster assignments. For example, if vertices $i$ and $j$ belong to clusters $\sigma$ and $\sigma^{\prime}$, respectively, they are connected with probability $\omega_{\sigma\sigma^{\prime}}$ ($\sigma, \sigma^{\prime} \in \{1, \dots, q\}$); matrix $\ket{\omega}$ is called the affinity matrix. For given $N$, $q$, $\ket{\gamma}$, and $\ket{\omega}$, we can generate random graph instances of the stochastic block model. In the case of the inverse problem, which is of interest to us in this paper, our goal is to infer the parameters $\ket{\gamma}$ and $\ket{\omega}$ as well as cluster assignments $\ket{\sigma}$ given a graph. The number of clusters $q$ is sometimes given as input; otherwise, it is determined by some model selection criterion. 
Throughout this paper, we treat $q$ as input and focus on sparse graphs, i.e., each element of $\ket{\omega}$ is scaled as $O(1/N)$ so that the average degree does not diverge as $N \to \infty$. 

\begin{figure}[t]
 \begin{center}
   \includegraphics[width=0.9 \columnwidth]{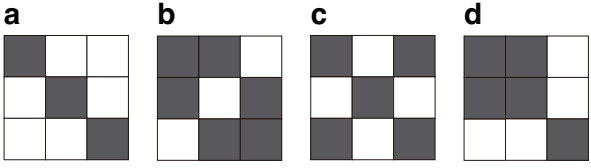}
 \end{center}
 \caption{
	Affinity matrices of various modular structures. The elements in gray have higher connection probabilities. 
  }
 \label{FigAffinityMatrices}
\end{figure}

While there exist many types of modular structures, the simplest and most studied case is the community structure as illustrated in Fig.~\ref{FigAffinityMatrices}(a); that is, the affinity matrix has large values for its diagonal elements, $\omega_{\sigma\sigma} = \omega_{\mathrm{in}}$, and small values for the remaining elements, $\omega_{\sigma\sigma^{\prime}} = \omega_{\mathrm{out}}$ ($\sigma \ne \sigma^{\prime}$). Although the elements of the affinity matrix can be arbitrary nonnegative numbers, we hereafter consider the case where they are either $\omega_{\mathrm{in}}$ or $\omega_{\mathrm{out}}$: that is, 
\begin{align}
\ket{\omega} = (\omega_{\mathrm{in}} - \omega_{\mathrm{out}}) W + \omega_{\mathrm{out}} \ket{1}\bra{1}, \label{AffinityMatrix}
\end{align}
where $W$ is an indicator matrix, where $W_{\sigma \sigma^{\prime}} = 1$ represents a densely connected cluster pair (which we refer to as a bicluster), $W_{\sigma \sigma^{\prime}} = 0$ represents a sparsely connected bicluster, and $\ket{1}$ is the column vector with all elements equal to unity. 
This random graph ensemble can be regarded as a restricted version of the stochastic block model, or a generalized version of the planted partition model \cite{Condon2001}. 

This affinity matrix contains the above community structure as a special case, and can express arbitrary modular patterns. Note that the indicator matrix $W$ can be regarded as a cluster-wise adjacency matrix, i.e., each planted cluster represents a coarse-grained vertex and a densely connected bicluster represents a bundled edge (a densely connected cluster constitutes a self-loop). We refer to the graph with adjacency matrix equal to $W$ as a \textit{module graph}. Note that some matrices represent the equivalent modular pattern; for example, Figs.~\ref{FigAffinityMatrices}(c) and \ref{FigAffinityMatrices}(d) differ only by permutation. The average degree $c$ of this stochastic block model is $c = N \bra{\gamma} \ket{\omega} \ket{\gamma}$. By defining the strength of the modular structure by $\epsilon \equiv \omega_{\mathrm{out}}/\omega_{\mathrm{in}}$, we can express elements $\omega_{\mathrm{in}}$ and $\omega_{\mathrm{out}}$ as 
\begin{align}
&\omega_{\mathrm{in}} = \frac{c}{N} \left[ (1-\epsilon) \bra{\gamma} W \ket{\gamma} + \epsilon \right]^{-1}, 
&\omega_{\mathrm{out}} = \epsilon \, \omega_{\mathrm{in}}. 
\end{align}

\section{Bayesian inference of the stochastic block model}\label{SBMinference}
We now consider the Bayesian inference of the modular structure using the stochastic block model. The prior probability $p(\ket{\sigma} \lvert \ket{\gamma})$ of cluster assignments is represented by a multinomial distribution of each planted cluster of fraction $\gamma_{\sigma}$, and the probability of independent and random connections between vertex pairs is represented by the product of Bernoulli distributions. Thus, the likelihood of the stochastic block model is 
\begin{align}
p(A, \ket{\sigma} \lvert \ket{\omega}, \ket{\gamma},q) 
&= p(A \lvert \ket{\sigma}, \ket{\omega}, \ket{\gamma}) p(\ket{\sigma} \lvert \ket{\gamma}) \notag\\
&= \prod_{i} \gamma_{\sigma_{i}} \prod_{i<j} \omega_{\sigma_{i}\sigma_{j}}^{A_{ij}}\left( 1 - \omega_{\sigma_{i}\sigma_{j}} \right)^{1-A_{ij}}. 
\end{align}
Using the affinity matrix of (\ref{AffinityMatrix}), its log-likelihood reads as
\begin{align}
& \log p(A, \ket{\sigma} \lvert \ket{\omega}, \ket{\gamma},q) 
= \sum_{i} \log \gamma_{\sigma_{i}} \notag\\ 
&\hspace{5pt} + \sum_{i<j} W_{\sigma_{i} \sigma_{j}} \left( A_{ij}\log\omega_{\mathrm{in}} + (1-A_{ij})\log(1-\omega_{\mathrm{in}}) \right) \notag\\
&\hspace{5pt} + \sum_{i<j} \left( 1 - W_{\sigma_{i} \sigma_{j}} \right) \left( A_{ij}\log\omega_{\mathrm{out}} + (1-A_{ij})\log(1-\omega_{\mathrm{out}}) \right). \label{SBMloglikelihood}
\end{align}

Our task is to evaluate the marginal probability distributions of the cluster assignments of vertices and to determine the values of parameters ($\ket{\gamma}$ and $\ket{\omega}$), in order to maximize the marginal log-likelihood 
\begin{align}
\log \sum_{\ket{\sigma}} p(A, \ket{\sigma} \lvert \ket{\gamma}, \ket{\omega},q). \label{MarginalLogLikelihood}
\end{align}
To this end, we employ the expectation-maximization (EM) algorithm, which does not maximize (\ref{MarginalLogLikelihood}) directly, but repeats the maximization of its lower bound until convergence: In the E-step, the posterior distribution of cluster assignments $\ket{\sigma}$ is estimated according to the given parameter estimates $(\ket{\gamma}, \ket{\omega})$. In the M-step, $(\ket{\gamma}, \ket{\omega})$ are updated to maximize the average of (\ref{SBMloglikelihood}) with respect to the posterior distribution determined in the E-step. 
While there are many other Bayesian inference methods \cite{Nowicki2001,daudin08,latouche12,PeixotoPRE2014MonteCarlo}, as we see below, the present method is suited for theoretical analysis. 

\subsection{Cluster inference and parameter learning}
Let $\psi^{i}_{\sigma}$ be the marginal probability of cluster $\sigma$ for vertex $i$ calculated in the E-step ($\sum_{\sigma} \psi^{i}_{\sigma} = 1$), and $\ket{\psi}^{i}$ be its row vector. Unfortunately, the exact computation of $\ket{\psi}^{i}$ is demanding. To avoid this computational burden, we use BP \cite{Decelle2011a,MezardMontanari2009}, which is justified for sparse graphs. Using tree approximation, the marginal probability $\ket{\psi}^{i}$ can be estimated as 
\begin{align}
\ket{\psi}^{i}
&= \frac{1}{Z^{i}} \ket{\gamma} \circ 
\prod_{k \in \partial i} \left[ \ket{1} + \overline{\omega}_{\mathrm{in}} \ket{\psi}^{k \to i} W \right] \circ 
\exp \left[ -\overline{\omega}_{\mathrm{in}} \omega_{\mathrm{out}} \sum_{\ell} \ket{\psi}^{\ell} W \right], \label{CompleteMarginal}
\end{align}
where $\ket{1}$ and $\ket{\psi}^{k \to i}$ are the $q$-dimensional unit row-vector and the marginal probability for vertex $k$ without the contribution from edge $(k, i)$, respectively. The latter is often referred to as the cavity bias. $\circ$ and $\partial i$ represent the element-wise product (Hadamard product) and the set of neighboring vertices of vertex $i$, respectively, and $Z^{i}$ is the normalization factor. 
We also define
\begin{align}
\overline{\omega}_{\mathrm{in}} 
&\equiv \frac{\omega_{\mathrm{in}} - \omega_{\mathrm{out}}}{\omega_{\mathrm{out}}} 
= \epsilon^{-1} - 1.
\end{align}

To obtain $\ket{\psi}^{i \to j}$, we compute the following iterative equation, i.e., the BP update equation. 
\begin{align}
\ket{\psi}^{i \to j} 
&= \frac{1}{Z^{i \to j}} \ket{\gamma} \circ 
\prod_{k \in \partial i \backslash j} \left[ \ket{1} + \overline{\omega}_{\mathrm{in}} \ket{\psi}^{k \to i} W \right] \notag\\
&\hspace{70pt} \circ \exp \left[ -\overline{\omega}_{\mathrm{in}} \omega_{\mathrm{out}} \sum_{\ell} \ket{\psi}^{\ell} W \right]. \label{GeneralBPequation1}
\end{align}
Analogously to (\ref{CompleteMarginal}), $Z^{i \to j}$ is the normalization factor. The BP update equation (\ref{GeneralBPequation1}) can be formally written as 
\begin{align}
\ket{\psi}^{i \to j} &= \mathcal{F}^{i \to j}\left[ \ket{\psi}^{k \to i} W, \ket{\psi}^{\ell} W \right], \label{GeneralBPequation1-2}
\end{align}
where $\mathcal{F}^{i \to j}$ is the non-linear operator representing the right-hand side of (\ref{GeneralBPequation1}). Note that $\ket{\psi}^{i \to j} = \mathcal{F}^{i \to j}\left[ \ket{\psi}^{k \to i}, \ket{\psi}^{\ell} \right]$ is essentially equivalent to the so-called mod-bp \cite{ZhangMoore2014} (without degree correction). If we consider cavity biases $\ket{\Psi}^{i \to j}$ of the transformed basis 
\begin{align}
\ket{\Psi}^{i \to j} &\equiv \ket{\psi}^{i \to j} W, 
\end{align}
its update equation is
\begin{align}
\ket{\Psi}^{i \to j} &= \mathcal{F}^{i \to j}\left[ \ket{\Psi}^{k \to i}, \ket{\Psi}^{\ell} \right] W. \label{GeneralBPequation2}
\end{align}
We can transform back to the original basis by operating $W^{-1}$ if it exists, or by operating $\mathcal{F}^{i \to j}$.

In the M-step, the parameter estimates ($\hat{\ket{\gamma}}$ and $\hat{\ket{\omega}}$) are updated as 
\begin{align}
\hat{\gamma}_{\sigma} &= \frac{1}{N}\sum_{i=1}^{N} \bracket{\delta_{\sigma \sigma_{i}}}, \label{hatgammaSBM}\\
\hat{\omega}_{\mathrm{in}} &= \frac{\sum_{i<j}A_{ij} \bracket{W_{\sigma_{i}\sigma_{j}}}}{\sum_{i<j}\bracket{W_{\sigma_{i}\sigma_{j}}}}, \label{hatomegainSBM}\\
\hat{\omega}_{\mathrm{out}} &= \frac{\sum_{i<j}A_{ij} \left( 1 - \bracket{W_{\sigma_{i}\sigma_{j}}}\right)}{\sum_{i<j}\left( 1 - \bracket{W_{\sigma_{i}\sigma_{j}}}\right)} \label{hatomegaoutSBM}, 
\end{align}
which can be readily obtained by the extremum conditions, where $\delta_{\sigma\sigma^{\prime}}$ is the Kronecker delta and $\bracket{\cdots} = \sum_{\ket{\sigma}} \cdots p(\ket{\sigma} \lvert \hat{\ket{\gamma}},\hat{\ket{\omega}},A)$ represents the average with respect to cluster assignments based on previous parameter estimates. 
Using the marginal probability estimates $\{\ket{\psi}^{i}\}$ and cavity biases $\{\ket{\psi}^{i \to j}\}$, we obtain $\bracket{\delta_{\sigma \sigma_{i}}} = \psi^{i}_{\sigma}$ and 
\begin{align}
\bracket{W_{\sigma_{i}\sigma_{j}}} 
&= \frac{\omega_{\mathrm{in}} \ket{\psi}^{i \to j} W \ket{\psi}^{j \to i \top}}{(\omega_{\mathrm{in}} - \omega_{\mathrm{out}}) \ket{\psi}^{i \to j} W \ket{\psi}^{j \to i \top} + \omega_{\mathrm{out}} }. \label{Westimate}
\end{align}
Assuming that cluster assignments are narrowly peaked \cite{ZhangMartinNewman2015}, we can approximate the denominator of (\ref{hatomegainSBM}) as 
\begin{align}
\sum_{i<j}\bracket{W_{\sigma_{i}\sigma_{j}}} 
\approx \frac{1}{2} \sum_{i,j} \ket{\psi}^{i} W \ket{\psi}^{j \top}. 
\end{align}
Note that we do not directly maximize (\ref{MarginalLogLikelihood}). Instead, by iteratively updating (\ref{GeneralBPequation1}) and (\ref{hatgammaSBM})--(\ref{hatomegaoutSBM}), the algorithm reaches a local extremum of the approximated marginal likelihood, or the negative Bethe free energy, which is a good estimate of (\ref{MarginalLogLikelihood}) when the graph is sparse and is exact when the graph is a tree.

\section{Detectability threshold}
We now analyze the detectability threshold for a given affinity matrix $W$. In the undetectable phase, BP converges to a trivial (uninformative) fixed point. When the graph reaches the detectable phase, the trivial fixed point becomes unstable, and BP converges to an informative fixed point instead. To see this stability, we first consider the propagation of perturbations on a vertex at the trivial fixed point. In the linear-response regime, it is dominated by the transfer matrix of (\ref{GeneralBPequation2}) 
\begin{align}
T_{\sigma^{\prime} \sigma}
= \frac{\delta \Psi^{i \to j}_{\sigma}}{\delta \Psi^{k \to i}_{\sigma^{\prime}}} 
&= \frac{\overline{\omega}_{\mathrm{in}}}{1 + \overline{\omega}_{\mathrm{in}} \Psi^{k \to i}_{\sigma^{\prime}}} \psi^{i \to j}_{\sigma^{\prime}} \left( W_{\sigma^{\prime} \sigma} - \Psi^{i \to j}_{\sigma} \right). \label{TransferMatrix}
\end{align}
We neglect the contribution due to $\overline{\omega}_{\mathrm{in}} \omega_{\mathrm{out}} \sum_{\ell} \Psi^{\ell}_{\tilde{\sigma}}$, because $\omega_{\mathrm{out}} = O(1/N)$. 

Although the effect of the perturbation of a single vertex may be vanishingly small at a distant vertex, if the effect from all connected vertices adds to $O(1)$, the trivial fixed point is unstable. Under tree approximation, this is achieved when $c \nu^{2} > 1$, where $\nu$ is the leading eigenvalue of the transfer matrix $T$; the equality condition yields the detectability threshold. Note that investigating the detectability threshold for an arbitrary structure is difficult because the trivial fixed point is not always known. In the following, hence, we analyze some solvable cases.

\begin{figure}[t]
 \begin{center}
   \includegraphics[width=0.88 \columnwidth]{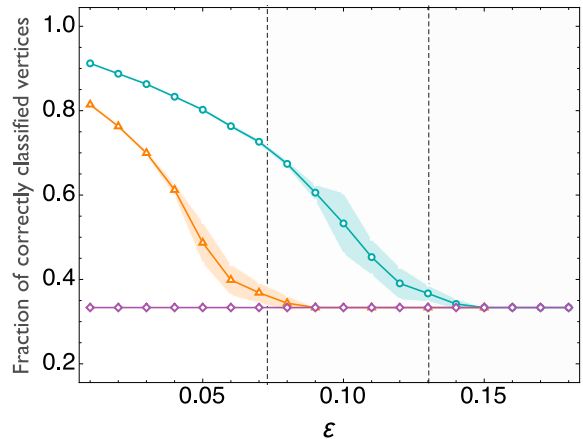}
 \end{center}
 \caption{
	(Color online) Fraction of correctly classified vertices for the structure of Fig.~\ref{FigAffinityMatrices}b. 
	The size of the graph is $N=30,000$, and each cluster is equal in size. 
	The connected diamonds (purple), triangles (orange), and circles (cyan) represent the results of the algorithm in Sec.~\ref{SBMinference} for the average degrees $c = 4, 5$, and $6$, respectively. 
	The dashed vertical lines are the detectability thresholds predicted in (\ref{RegularDetectability}) for $c = 5$ and $6$. 
	The shadows represent the standard deviations of $10$ samples. 
}
 \label{DetectabilityW6}
\end{figure}

\subsection{A solvable case}
Let us consider the case where a fraction of clusters is equal in size, i.e., $\gamma_{\sigma} = 1/q$ for any $\sigma$, and the average degree of each cluster is also equal. That is, 
\begin{align}
&\sum_{\sigma^{\prime}} W_{\sigma \sigma^{\prime}} = a  & (a = \text{const.}) \label{RegularCondition}
\end{align}
for any $\sigma$. In other words, the module graph constitutes a regular graph. This is also assumed in Ref.~\cite{Decelle2011a}. In this case, the factorized state, i.e., $\psi^{i \to j}_{\sigma} = 1/q$ for any $i \to j$ and $\sigma$, is the trivial BP fixed point. Therefore, the transfer matrix $T$ at this fixed point is
\begin{align}
T = \frac{\overline{\omega}_{\mathrm{in}}}{ q + a \overline{\omega}_{\mathrm{in}} } \left( W - \frac{a}{q}\ket{1}\bra{1} \right). 
\end{align}
Because $\ket{1}/\sqrt{q}$ is the leading eigenvector of $W$ with eigenvalue $a$, $\nu$ can be written as 
\begin{align}
\nu = \frac{\overline{\omega}_{\mathrm{in}}}{ q + a \overline{\omega}_{\mathrm{in}} } \lambda_{2}, 
\end{align}
where $\lambda_{2}$ is the second leading eigenvalue of $W$ in magnitude. Thus, in terms of $\epsilon$, the detectability threshold is given by 
\begin{align}
\epsilon^{\ast} = \frac{|\lambda_{2}|\sqrt{c} - a}{|\lambda_{2}|\sqrt{c} - a + q}. \label{RegularDetectability}
\end{align}

The stochastic block model with a community structure has $a=1$ and $\lambda_{2} = 1$, which reproduces a previously known result \cite{Decelle2011a}. The threshold (\ref{RegularDetectability}) indicates that as the number of densely connected clusters increases, the difficulty in inferring the structure also increases. In particular, when $c < (a/\lambda_{2})^{2}$, it is statistically impossible to infer the planted structure better than chance for any $\epsilon$. This behavior is shown in Fig.~\ref{DetectabilityW6}; when $c=4$, no signal is retrieved even when the noise $\epsilon$ is (almost) zero. 

The $\lambda_{2}$-dependency of the module graph in (\ref{RegularDetectability}) is another notable feature. 
For graph $G$, the second eigenvalue $\lambda_{2}$ of an adjacency matrix is bounded from below and above by the (normalized) edge expansion $h(G)$ as 
\begin{align}
1 - 2 h(G) \le \lambda_{2} \le 1 - \frac{h(G)^{2}}{2}, \label{CheegerInequality}
\end{align}
which is known as Cheeger's inequality \cite{chung1996spectral}. The edge expansion $h(G)$ is a measure of a sparse cut, defined by 
\begin{align}
h(G) = \min_{S} \frac{\lvert E(S, V \backslash S)\rvert}{a \min\{\lvert S \rvert, \lvert V \backslash S \rvert\}}, 
\end{align}
where $S$ is a subset of vertex set $V$ of the graph, and $\lvert E(S, V \backslash S)\rvert$ is the number of edges between sets $S$ and $V \backslash S$. The inequality (\ref{CheegerInequality}) indicates that the module graph with no satisfactory sparse cut [large $h(G)$] tends to have a small value of $\lambda_{2}$: that is, the planted structure is difficult to infer. Put another way, if the graph has a strong hierarchical modular structure \cite{FootnoteHierarchicalStructure}, its inference tends to be relatively easy. Note also that as long as the second eigenvalue is strictly positive, the detectability threshold is always positive for a sufficiently large average degree. 

One might think that a different detectability threshold can be obtained if we instead use the flipped indicator matrix $\widetilde{W} = \ket{1}\bra{1} - W$ to parametrize noise strength as $\tilde{\epsilon} \equiv \epsilon^{-1}$, even though the structure to infer is the same. However, one can straightforwardly confirm that this treatment also yields threshold $\tilde{\epsilon}^{\ast}$ equal to (\ref{RegularDetectability}). 

\subsection{Another solvable case}
In the case where the factorized state is not a trivial BP fixed point, the calculation of the detectability threshold is difficult. Although it is rather a toy model example, there is another case where we can obtain the analytical expression for it. 

\begin{figure}[t]
 \begin{center}
   \includegraphics[width=0.88 \columnwidth]{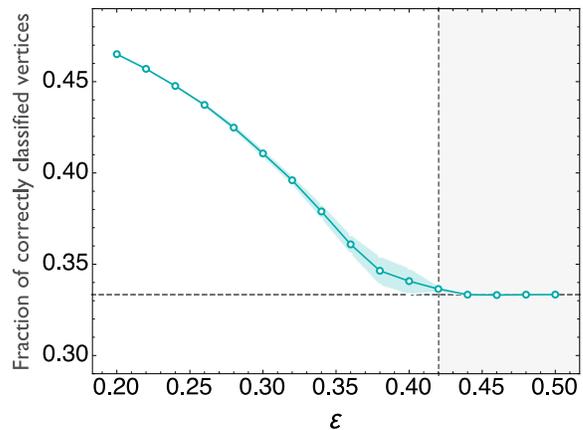}
 \end{center}
 \caption{
	(Color online) Fraction of correctly classified vertices for the structure of Figs.~\ref{FigAffinityMatrices}(c) with error bars. 
	The dashed vertical and horizontal lines represent the estimate of the detectability threshold (\ref{DetectabilityThresholdW8}) and $1/3$, respectively. 
	The size of the graph is $N=30,000$ with average degree $c=6$ and each cluster has the same size. 
	The shadow represents the standard deviation of $10$ samples. 
}
 \label{DetectabilityW8}
\end{figure}

Let $W$ be a matrix whose linearly independent columns are orthogonal to one another, e.g., Fig.~\ref{FigAffinityMatrices}(c). 
We set the prior distribution $\ket{\gamma}$ so that $\ket{\gamma}W \propto \ket{1}^{\top}$, and keep it fixed, i.e., we skip (\ref{hatgammaSBM}); for the structure in Fig.~\ref{FigAffinityMatrices}(c), we set $\ket{\gamma} = (1/4, 1/2, 1/4)$, although the fractions of the planted clusters do not have this ratio. In this case, the factorized fixed point is a BP fixed point. For this example, the transfer matrix (\ref{TransferMatrix}) reads 
\begin{align}
T = 
\frac{\overline{\omega}_{\mathrm{in}}}{4(2 + \overline{\omega}_{\mathrm{in}})}
\begin{pmatrix}
1 & -1 & 1 \\
-2 & 2 & -2 \\
1 & -1 & 1 
\end{pmatrix} \label{Texample}
\end{align}
and the leading eigenvalue is $\nu = \overline{\omega}_{\mathrm{in}}(2 + \overline{\omega}_{\mathrm{in}})^{-1}$. The corresponding detectability threshold is 
\begin{align}
\epsilon^{\ast} = \frac{\sqrt{c}-1}{\sqrt{c}+1}. \label{DetectabilityThresholdW8}
\end{align}
This threshold was compared with the numerical experiment in Fig.~\ref{DetectabilityW8}.

\section{Summary and Discussion}
In this paper, we analyzed the detectability thresholds of general modular structures in the restricted graph ensembles. 
Although our results do not cover arbitrary structures, our solvable case analyses provide deeper insight into the nature of detectability. We showed that some structures are statistically impossible to infer (using BP in Sec.~\ref{SBMinference}), no matter how small the noise $\epsilon$ is. We also revealed that detectability transition is connected to the hierarchical structure of clusters. 
Our results are not rigorous and may differ from the information-theoretic limits. 
Also, when the number of clusters is large, there often exists another phase called the \textit{hard phase} \cite{Decelle2011a}. These points are left as open questions for future research.

\section*{acknowledgments}
The authors thank Jean-Gabriel Young for useful comments. 
This work was supported by JSPS KAKENHI No. 26011023 (T.K.) and No. 25120013 (Y.K.).

\bibliographystyle{apsrev}
\bibliography{bib-generalSBM}

\begin{thebibliography}{40}
\expandafter\ifx\csname natexlab\endcsname\relax\def\natexlab#1{#1}\fi
\expandafter\ifx\csname bibnamefont\endcsname\relax
  \def\bibnamefont#1{#1}\fi
\expandafter\ifx\csname bibfnamefont\endcsname\relax
  \def\bibfnamefont#1{#1}\fi
\expandafter\ifx\csname citenamefont\endcsname\relax
  \def\citenamefont#1{#1}\fi
\expandafter\ifx\csname url\endcsname\relax
  \def\url#1{\texttt{#1}}\fi
\expandafter\ifx\csname urlprefix\endcsname\relax\def\urlprefix{URL }\fi
\providecommand{\bibinfo}[2]{#2}
\providecommand{\eprint}[2][]{\url{#2}}

\bibitem[{\citenamefont{Girvan and Newman}(2002)}]{GirvanNewman2002}
\bibinfo{author}{\bibfnamefont{M.}~\bibnamefont{Girvan}} \bibnamefont{and}
  \bibinfo{author}{\bibfnamefont{M.~E.~J.} \bibnamefont{Newman}},
  \bibinfo{journal}{Proc. Natl. Acad. Sci. U.S.A.}
  \textbf{\bibinfo{volume}{99}}, \bibinfo{pages}{7821} (\bibinfo{year}{2002}).

\bibitem[{\citenamefont{Shi and Malik}(2000)}]{ShiMalik2000}
\bibinfo{author}{\bibfnamefont{J.}~\bibnamefont{Shi}} \bibnamefont{and}
  \bibinfo{author}{\bibfnamefont{J.}~\bibnamefont{Malik}},
  \bibinfo{journal}{IEEE Trans. Pattern Anal. and Machine Intel.}
  \textbf{\bibinfo{volume}{22}}, \bibinfo{pages}{888} (\bibinfo{year}{2000}).

\bibitem[{\citenamefont{Leskovec et~al.}(2009)\citenamefont{Leskovec, Lang,
  Anirban, and Mahoney}}]{Leskovec2009}
\bibinfo{author}{\bibfnamefont{J.}~\bibnamefont{Leskovec}},
  \bibinfo{author}{\bibfnamefont{K.~J.} \bibnamefont{Lang}},
  \bibinfo{author}{\bibfnamefont{D.}~\bibnamefont{Anirban}}, \bibnamefont{and}
  \bibinfo{author}{\bibfnamefont{M.~W.} \bibnamefont{Mahoney}},
  \bibinfo{journal}{Internet Math.} \textbf{\bibinfo{volume}{6}},
  \bibinfo{pages}{29} (\bibinfo{year}{2009}).

\bibitem[{\citenamefont{Fortunato}(2010)}]{Fortunato201075}
\bibinfo{author}{\bibfnamefont{S.}~\bibnamefont{Fortunato}},
  \bibinfo{journal}{Phys. Rep.} \textbf{\bibinfo{volume}{486}},
  \bibinfo{pages}{75} (\bibinfo{year}{2010}).

\bibitem[{\citenamefont{Leger et~al.}(2014)\citenamefont{Leger, Vacher, and
  Daudin}}]{Leger2013}
\bibinfo{author}{\bibfnamefont{J.-B.} \bibnamefont{Leger}},
  \bibinfo{author}{\bibfnamefont{C.}~\bibnamefont{Vacher}}, \bibnamefont{and}
  \bibinfo{author}{\bibfnamefont{J.-J.} \bibnamefont{Daudin}},
  \bibinfo{journal}{Stat. Comp.} \textbf{\bibinfo{volume}{24}},
  \bibinfo{pages}{675} (\bibinfo{year}{2014}).

\bibitem[{\citenamefont{Holland et~al.}(1983)\citenamefont{Holland, Laskey, and
  Leinhardt}}]{holland1983stochastic}
\bibinfo{author}{\bibfnamefont{P.~W.} \bibnamefont{Holland}},
  \bibinfo{author}{\bibfnamefont{K.~B.} \bibnamefont{Laskey}},
  \bibnamefont{and}
  \bibinfo{author}{\bibfnamefont{S.}~\bibnamefont{Leinhardt}},
  \bibinfo{journal}{Soc. Netw.} \textbf{\bibinfo{volume}{5}},
  \bibinfo{pages}{109} (\bibinfo{year}{1983}).

\bibitem[{\citenamefont{Reichardt and Leone}(2008)}]{Reichardt2008}
\bibinfo{author}{\bibfnamefont{J.}~\bibnamefont{Reichardt}} \bibnamefont{and}
  \bibinfo{author}{\bibfnamefont{M.}~\bibnamefont{Leone}},
  \bibinfo{journal}{Phys. Rev. Lett.} \textbf{\bibinfo{volume}{101}},
  \bibinfo{pages}{078701} (\bibinfo{year}{2008}).

\bibitem[{\citenamefont{Nadakuditi and Newman}(2012)}]{Nadakuditi2012}
\bibinfo{author}{\bibfnamefont{R.~R.} \bibnamefont{Nadakuditi}}
  \bibnamefont{and} \bibinfo{author}{\bibfnamefont{M.~E.~J.}
  \bibnamefont{Newman}}, \bibinfo{journal}{Phys. Rev. Lett.}
  \textbf{\bibinfo{volume}{108}}, \bibinfo{pages}{188701}
  (\bibinfo{year}{2012}).

\bibitem[{\citenamefont{Krzakala et~al.}(2013)\citenamefont{Krzakala, Moore,
  Mossel, Neeman, Sly, Zdeborov\'{a}, and Zhang}}]{Krzakala2013}
\bibinfo{author}{\bibfnamefont{F.}~\bibnamefont{Krzakala}},
  \bibinfo{author}{\bibfnamefont{C.}~\bibnamefont{Moore}},
  \bibinfo{author}{\bibfnamefont{E.}~\bibnamefont{Mossel}},
  \bibinfo{author}{\bibfnamefont{J.}~\bibnamefont{Neeman}},
  \bibinfo{author}{\bibfnamefont{A.}~\bibnamefont{Sly}},
  \bibinfo{author}{\bibfnamefont{L.}~\bibnamefont{Zdeborov\'{a}}},
  \bibnamefont{and} \bibinfo{author}{\bibfnamefont{P.}~\bibnamefont{Zhang}},
  \bibinfo{journal}{Proc. Natl. Acad. Sci. U.S.A.}
  \textbf{\bibinfo{volume}{110}}, \bibinfo{pages}{20935}
  (\bibinfo{year}{2013}).

\bibitem[{\citenamefont{Kawamoto and
  Kabashima}(2015{\natexlab{a}})}]{KawamotoKabashimaPRE2015}
\bibinfo{author}{\bibfnamefont{T.}~\bibnamefont{Kawamoto}} \bibnamefont{and}
  \bibinfo{author}{\bibfnamefont{Y.}~\bibnamefont{Kabashima}},
  \bibinfo{journal}{Phys. Rev. E} \textbf{\bibinfo{volume}{91}},
  \bibinfo{pages}{062803} (\bibinfo{year}{2015}{\natexlab{a}}).

\bibitem[{\citenamefont{Kawamoto and
  Kabashima}(2015{\natexlab{b}})}]{KawamotoKabashimaEPL2015}
\bibinfo{author}{\bibfnamefont{T.}~\bibnamefont{Kawamoto}} \bibnamefont{and}
  \bibinfo{author}{\bibfnamefont{Y.}~\bibnamefont{Kabashima}},
  \bibinfo{journal}{Eur. Phys. Lett.} \textbf{\bibinfo{volume}{112}},
  \bibinfo{pages}{40007} (\bibinfo{year}{2015}{\natexlab{b}}).

\bibitem[{\citenamefont{Decelle
  et~al.}(2011{\natexlab{a}})\citenamefont{Decelle, Krzakala, Moore, and
  Zdeborov\'{a}}}]{Decelle2011}
\bibinfo{author}{\bibfnamefont{A.}~\bibnamefont{Decelle}},
  \bibinfo{author}{\bibfnamefont{F.}~\bibnamefont{Krzakala}},
  \bibinfo{author}{\bibfnamefont{C.}~\bibnamefont{Moore}}, \bibnamefont{and}
  \bibinfo{author}{\bibfnamefont{L.}~\bibnamefont{Zdeborov\'{a}}},
  \bibinfo{journal}{Phys. Rev. Lett.} \textbf{\bibinfo{volume}{107}},
  \bibinfo{pages}{065701} (\bibinfo{year}{2011}{\natexlab{a}}).

\bibitem[{\citenamefont{Decelle
  et~al.}(2011{\natexlab{b}})\citenamefont{Decelle, Krzakala, Moore, and
  Zdeborov\'{a}}}]{Decelle2011a}
\bibinfo{author}{\bibfnamefont{A.}~\bibnamefont{Decelle}},
  \bibinfo{author}{\bibfnamefont{F.}~\bibnamefont{Krzakala}},
  \bibinfo{author}{\bibfnamefont{C.}~\bibnamefont{Moore}}, \bibnamefont{and}
  \bibinfo{author}{\bibfnamefont{L.}~\bibnamefont{Zdeborov\'{a}}},
  \bibinfo{journal}{Phys. Rev. E} \textbf{\bibinfo{volume}{84}},
  \bibinfo{pages}{066106} (\bibinfo{year}{2011}{\natexlab{b}}).

\bibitem[{\citenamefont{Radicchi}(2013)}]{Radicchi2013}
\bibinfo{author}{\bibfnamefont{F.}~\bibnamefont{Radicchi}},
  \bibinfo{journal}{Phys. Rev. E} \textbf{\bibinfo{volume}{88}},
  \bibinfo{pages}{010801} (\bibinfo{year}{2013}).

\bibitem[{\citenamefont{Radicchi}(2014)}]{Radicchi2014}
\bibinfo{author}{\bibfnamefont{F.}~\bibnamefont{Radicchi}},
  \bibinfo{journal}{Eur. Phys. Lett.} \textbf{\bibinfo{volume}{106}},
  \bibinfo{pages}{38001} (\bibinfo{year}{2014}).

\bibitem[{\citenamefont{Hu et~al.}(2012)\citenamefont{Hu, Ronhovde, and
  Nussinov}}]{Hu2012}
\bibinfo{author}{\bibfnamefont{D.}~\bibnamefont{Hu}},
  \bibinfo{author}{\bibfnamefont{P.}~\bibnamefont{Ronhovde}}, \bibnamefont{and}
  \bibinfo{author}{\bibfnamefont{Z.}~\bibnamefont{Nussinov}},
  \bibinfo{journal}{Philos. Mag.} \textbf{\bibinfo{volume}{92}},
  \bibinfo{pages}{406} (\bibinfo{year}{2012}).

\bibitem[{\citenamefont{Ronhovde et~al.}(2012)\citenamefont{Ronhovde, Hu, and
  Nussinov}}]{Ronhovde2012}
\bibinfo{author}{\bibfnamefont{P.}~\bibnamefont{Ronhovde}},
  \bibinfo{author}{\bibfnamefont{D.}~\bibnamefont{Hu}}, \bibnamefont{and}
  \bibinfo{author}{\bibfnamefont{Z.}~\bibnamefont{Nussinov}},
  \bibinfo{journal}{Eur. Phys. Lett.} \textbf{\bibinfo{volume}{99}},
  \bibinfo{pages}{38006} (\bibinfo{year}{2012}).

\bibitem[{\citenamefont{Steeg et~al.}(2014)\citenamefont{Steeg, Moore,
  Galstyan, and Allahverdyan}}]{VerSteeg2014}
\bibinfo{author}{\bibfnamefont{G.~V.} \bibnamefont{Steeg}},
  \bibinfo{author}{\bibfnamefont{C.}~\bibnamefont{Moore}},
  \bibinfo{author}{\bibfnamefont{A.}~\bibnamefont{Galstyan}}, \bibnamefont{and}
  \bibinfo{author}{\bibfnamefont{A.}~\bibnamefont{Allahverdyan}},
  \bibinfo{journal}{Eur. Phys. Lett.} \textbf{\bibinfo{volume}{106}},
  \bibinfo{pages}{48004} (\bibinfo{year}{2014}).

\bibitem[{\citenamefont{Zhang et~al.}(2014)\citenamefont{Zhang, Moore, and
  Zdeborov\'a}}]{ZhangPRE2014}
\bibinfo{author}{\bibfnamefont{P.}~\bibnamefont{Zhang}},
  \bibinfo{author}{\bibfnamefont{C.}~\bibnamefont{Moore}}, \bibnamefont{and}
  \bibinfo{author}{\bibfnamefont{L.}~\bibnamefont{Zdeborov\'a}},
  \bibinfo{journal}{Phys. Rev. E} \textbf{\bibinfo{volume}{90}},
  \bibinfo{pages}{052802} (\bibinfo{year}{2014}).

\bibitem[{\citenamefont{Ghasemian et~al.}(2016)\citenamefont{Ghasemian, Zhang,
  Clauset, Moore, and Peel}}]{Ghasemian2016}
\bibinfo{author}{\bibfnamefont{A.}~\bibnamefont{Ghasemian}},
  \bibinfo{author}{\bibfnamefont{P.}~\bibnamefont{Zhang}},
  \bibinfo{author}{\bibfnamefont{A.}~\bibnamefont{Clauset}},
  \bibinfo{author}{\bibfnamefont{C.}~\bibnamefont{Moore}}, \bibnamefont{and}
  \bibinfo{author}{\bibfnamefont{L.}~\bibnamefont{Peel}},
  \bibinfo{journal}{Phys. Rev. X} \textbf{\bibinfo{volume}{6}},
  \bibinfo{pages}{031005} (\bibinfo{year}{2016}).

\bibitem[{\citenamefont{Mossel et~al.}(2014)\citenamefont{Mossel, Neeman, and
  Sly}}]{Mossel2014}
\bibinfo{author}{\bibfnamefont{E.}~\bibnamefont{Mossel}},
  \bibinfo{author}{\bibfnamefont{J.}~\bibnamefont{Neeman}}, \bibnamefont{and}
  \bibinfo{author}{\bibfnamefont{A.}~\bibnamefont{Sly}},
  \bibinfo{journal}{Probab. Theory Relat. Fields} pp. \bibinfo{pages}{1--31}
  (\bibinfo{year}{2014}).

\bibitem[{\citenamefont{Massouli{\'e}}(2014)}]{Massoulie2014}
\bibinfo{author}{\bibfnamefont{L.}~\bibnamefont{Massouli{\'e}}}, in
  \emph{\bibinfo{booktitle}{Proceedings of the 46th Annual ACM Symposium on
  Theory of Computing}} (\bibinfo{publisher}{ACM}, \bibinfo{address}{New York},
  \bibinfo{year}{2014}), STOC '14, pp. \bibinfo{pages}{694--703}.

\bibitem[{\citenamefont{Banks et~al.}(2016)\citenamefont{Banks, Moore, Neeman,
  and Netrapalli}}]{banks2016information}
\bibinfo{author}{\bibfnamefont{J.}~\bibnamefont{Banks}},
  \bibinfo{author}{\bibfnamefont{C.}~\bibnamefont{Moore}},
  \bibinfo{author}{\bibfnamefont{J.}~\bibnamefont{Neeman}}, \bibnamefont{and}
  \bibinfo{author}{\bibfnamefont{P.}~\bibnamefont{Netrapalli}}, in
  \emph{\bibinfo{booktitle}{29th Annual Conference on Learning Theory}}
  (\bibinfo{year}{2016}), pp. \bibinfo{pages}{383--416}.

\bibitem[{\citenamefont{Condon and Karp}(2001)}]{Condon2001}
\bibinfo{author}{\bibfnamefont{A.}~\bibnamefont{Condon}} \bibnamefont{and}
  \bibinfo{author}{\bibfnamefont{R.~M.} \bibnamefont{Karp}},
  \bibinfo{journal}{Random Struct. Algorithms} \textbf{\bibinfo{volume}{18}},
  \bibinfo{pages}{116} (\bibinfo{year}{2001}).

\bibitem[{\citenamefont{Onsj{\"o} and Watanabe}(2006)}]{Onsjo2006}
\bibinfo{author}{\bibfnamefont{M.}~\bibnamefont{Onsj{\"o}}} \bibnamefont{and}
  \bibinfo{author}{\bibfnamefont{O.}~\bibnamefont{Watanabe}}, in
  \emph{\bibinfo{booktitle}{Algorithms and Computation}}
  (\bibinfo{publisher}{Springer, New York}, \bibinfo{year}{2006}), pp.
  \bibinfo{pages}{507--516}.

\bibitem[{\citenamefont{Bickel and Chen}(2009)}]{BickelChenPNAS2009}
\bibinfo{author}{\bibfnamefont{P.~J.} \bibnamefont{Bickel}} \bibnamefont{and}
  \bibinfo{author}{\bibfnamefont{A.}~\bibnamefont{Chen}},
  \bibinfo{journal}{Proc. Natl. Acad. Sci. U.S.A.}
  \textbf{\bibinfo{volume}{106}}, \bibinfo{pages}{21068}
  (\bibinfo{year}{2009}).

\bibitem[{\citenamefont{Rohe et~al.}(2011)\citenamefont{Rohe, Chatterjee, and
  Yu}}]{Rohe2011}
\bibinfo{author}{\bibfnamefont{K.}~\bibnamefont{Rohe}},
  \bibinfo{author}{\bibfnamefont{S.}~\bibnamefont{Chatterjee}},
  \bibnamefont{and} \bibinfo{author}{\bibfnamefont{B.}~\bibnamefont{Yu}},
  \bibinfo{journal}{Ann. Stat.} \textbf{\bibinfo{volume}{39}},
  \bibinfo{pages}{1878} (\bibinfo{year}{2011}).

\bibitem[{\citenamefont{Yun and Proutiere}(2014)}]{yun2014community}
\bibinfo{author}{\bibfnamefont{S.-Y.} \bibnamefont{Yun}} \bibnamefont{and}
  \bibinfo{author}{\bibfnamefont{A.}~\bibnamefont{Proutiere}}, in
  \emph{\bibinfo{booktitle}{COLT}} (\bibinfo{year}{2014}), pp.
  \bibinfo{pages}{138--175}.

\bibitem[{\citenamefont{Abbe and Sandon}(2015{\natexlab{a}})}]{AbbeFOCS2015}
\bibinfo{author}{\bibfnamefont{E.}~\bibnamefont{Abbe}} \bibnamefont{and}
  \bibinfo{author}{\bibfnamefont{C.}~\bibnamefont{Sandon}}, in
  \emph{\bibinfo{booktitle}{2015 IEEE 56th Annual Symposium on Foundations of
  Computer Science (FOCS)}} (\bibinfo{year}{2015}{\natexlab{a}}), pp.
  \bibinfo{pages}{670--688}.

\bibitem[{\citenamefont{Abbe and Sandon}(2015{\natexlab{b}})}]{AbbeNIPS2015}
\bibinfo{author}{\bibfnamefont{E.}~\bibnamefont{Abbe}} \bibnamefont{and}
  \bibinfo{author}{\bibfnamefont{C.}~\bibnamefont{Sandon}}, in
  \emph{\bibinfo{booktitle}{Advances in Neural Information Processing Systems
  28}}, edited by \bibinfo{editor}{\bibfnamefont{C.}~\bibnamefont{Cortes}},
  \bibinfo{editor}{\bibfnamefont{N.~D.} \bibnamefont{Lawrence}},
  \bibinfo{editor}{\bibfnamefont{D.~D.} \bibnamefont{Lee}},
  \bibinfo{editor}{\bibfnamefont{M.}~\bibnamefont{Sugiyama}}, \bibnamefont{and}
  \bibinfo{editor}{\bibfnamefont{R.}~\bibnamefont{Garnett}}
  (\bibinfo{publisher}{Curran Associates}, \bibinfo{year}{2015}{\natexlab{b}}),
  pp. \bibinfo{pages}{676--684}.

\bibitem[{Foo({\natexlab{a}})}]{FootnoteAbbe}
\bibinfo{howpublished}{To the best of our knowledge,
  Refs.~\cite{AbbeFOCS2015,AbbeNIPS2015} are the only exceptions, in which the
  \textit{recovery} problem of the stochastic block model with general modular
  structure is considered.}

\bibitem[{\citenamefont{Nowicki and Snijders}(2001)}]{Nowicki2001}
\bibinfo{author}{\bibfnamefont{K.}~\bibnamefont{Nowicki}} \bibnamefont{and}
  \bibinfo{author}{\bibfnamefont{T.~A.~B.} \bibnamefont{Snijders}},
  \bibinfo{journal}{Journal of the American Statistical Association}
  \textbf{\bibinfo{volume}{96}}, \bibinfo{pages}{1077} (\bibinfo{year}{2001}).

\bibitem[{\citenamefont{Daudin et~al.}(2008)\citenamefont{Daudin, Picard, and
  Robin}}]{daudin08}
\bibinfo{author}{\bibfnamefont{J.~J.} \bibnamefont{Daudin}},
  \bibinfo{author}{\bibfnamefont{F.}~\bibnamefont{Picard}}, \bibnamefont{and}
  \bibinfo{author}{\bibfnamefont{S.}~\bibnamefont{Robin}},
  \bibinfo{journal}{Stat. Comp.} \textbf{\bibinfo{volume}{18}},
  \bibinfo{pages}{173} (\bibinfo{year}{2008}).

\bibitem[{\citenamefont{Latouche et~al.}(2012)\citenamefont{Latouche,
  Birmel\'{e}, and Ambroise}}]{latouche12}
\bibinfo{author}{\bibfnamefont{P.}~\bibnamefont{Latouche}},
  \bibinfo{author}{\bibfnamefont{E.}~\bibnamefont{Birmel\'{e}}},
  \bibnamefont{and} \bibinfo{author}{\bibfnamefont{C.}~\bibnamefont{Ambroise}},
  \bibinfo{journal}{Stat. Model.} \textbf{\bibinfo{volume}{12}},
  \bibinfo{pages}{93} (\bibinfo{year}{2012}).

\bibitem[{\citenamefont{Peixoto}(2014)}]{PeixotoPRE2014MonteCarlo}
\bibinfo{author}{\bibfnamefont{T.~P.} \bibnamefont{Peixoto}},
  \bibinfo{journal}{Phys. Rev. E} \textbf{\bibinfo{volume}{89}},
  \bibinfo{pages}{012804} (\bibinfo{year}{2014}).

\bibitem[{\citenamefont{M\'ezard and Montanari}(2009)}]{MezardMontanari2009}
\bibinfo{author}{\bibfnamefont{M.}~\bibnamefont{M\'ezard}} \bibnamefont{and}
  \bibinfo{author}{\bibfnamefont{A.}~\bibnamefont{Montanari}},
  \emph{\bibinfo{title}{Information, Physics, and Computation}}
  (\bibinfo{publisher}{Oxford University Press, Oxford}, \bibinfo{year}{2009}).

\bibitem[{\citenamefont{Zhang and Moore}(2014)}]{ZhangMoore2014}
\bibinfo{author}{\bibfnamefont{P.}~\bibnamefont{Zhang}} \bibnamefont{and}
  \bibinfo{author}{\bibfnamefont{C.}~\bibnamefont{Moore}},
  \bibinfo{journal}{Proc. Natl. Acad. Sci. U.S.A.}
  \textbf{\bibinfo{volume}{111}}, \bibinfo{pages}{18144}
  (\bibinfo{year}{2014}).

\bibitem[{\citenamefont{Zhang et~al.}(2015)\citenamefont{Zhang, Martin, and
  Newman}}]{ZhangMartinNewman2015}
\bibinfo{author}{\bibfnamefont{X.}~\bibnamefont{Zhang}},
  \bibinfo{author}{\bibfnamefont{T.}~\bibnamefont{Martin}}, \bibnamefont{and}
  \bibinfo{author}{\bibfnamefont{M.~E.~J.} \bibnamefont{Newman}},
  \bibinfo{journal}{Phys. Rev. E} \textbf{\bibinfo{volume}{91}},
  \bibinfo{pages}{032803} (\bibinfo{year}{2015}).

\bibitem[{\citenamefont{Chung}(1997)}]{chung1996spectral}
\bibinfo{author}{\bibfnamefont{F.~R.~K.} \bibnamefont{Chung}},
  \emph{\bibinfo{title}{Spectral Graph Theory (CBMS Regional Conference Series
  in Mathematics, No. 92)}} (\bibinfo{publisher}{American Mathematical Society,
  Providence, RI}, \bibinfo{year}{1997}).

\bibitem[{Foo({\natexlab{b}})}]{FootnoteHierarchicalStructure}
\bibinfo{howpublished}{Note that the graph $G$ here is the module graph. Thus,
  a large value of the second eigenvalue implies that the clusters constitute a
  (higher order) modular structure, and this is what we mean by the
  hierarchical structure.}

\end{thebibliography}

\end{document}